\documentclass{iaus}
\usepackage{graphicx}

\def\ftdm{flux transport dynamo model}
\def\tl{tachocline}

\def\ftlm{fast tachocline model}

\title[Flux Transport Dynamo and Fast Tachocline Scenario]%here short title %%
{Flux Transport Dynamo coupled with a Fast Tachocline Scenario}
\author[Bidya Binay Karak \& Kristof Petrovay]   %% give here short author list %%
{Bidya Binay Karak$^1$
\and Kristof Petrovay$^2$}
\affiliation{$^1$Department of Physics, Indian Institute of Science, Bangalore 560012, India\\email: {\tt bidya\_karak@physics.iisc.ernet.in}
$^2$E\"otv\"os~University,~Department~of~Astronomy,~Budapest,~Hungary}
\pubyear{2013}
\volume{294}  %% insert here IAU Symposium No.
\pagerange{119--126}
\jname{Solar and Astrophysical Dynamos and Magnetic Activity}
\editors{A.G. Kosovichev, E.M. de Gouveia Dal Pino, \& Y.Yan, eds.}
\begin{document}
\maketitle
\begin{abstract}
The tachocline is important in the solar dynamo for the generation and the storage of the 
magnetic fields. A most plausible explanation for the confinement of the tachocline 
is given by the fast tachocline model in which the tachocline is confined by the 
anisotropic momentum transfer by the Maxwell stress of the dynamo generated 
magnetic fields. We employ a flux transport dynamo model coupled with the simple feedback formula of this
fast tachocline model which basically relates the thickness of the tachocline to the Maxwell stress.
We find that this nonlinear coupling not only produces a stable solar-like dynamo solution 
but also a significant latitudinal variation in the tachocline thickness which is in 
agreement with the observations.
\end{abstract}
\firstsection % if your document starts with a section,

\section{Introduction}
The tachocline is a thin layer (of radial extent $< 20$~Mm) located at the base of the solar 
convection zone where the rotation changes from differential to the rigid rotation 
(e.g., Charbonneau {\it et al.} 1999). This layer is important in dynamo models
for the generation and the storage of the toroidal fields. However the thinness of this layer 
made the confinement of the tachocline an intriguing problem. Spiegel \& Zahn (1992) invoked 
strong anisotropic turbulent viscosity in the horizontal direction for the confinement of the 
tachocline. However several authors (e.g., Rudiger \& Kitchatinov, 1997; Gough \& McIntyre 1998) 
realized this purely hydrodynamical model to be inappropriate and proposed an 
alternative mechanism for the angular momentum transport. They have shown that a strongly anisotropic angular momentum transport 
is possible by invoking a weak fossil magnetic field in the radiative zone. This so-called {\it slow 
tachocline} model was later found to be questionable (e.g., Brun \& Zahn, 2006; 
Strugarek, Brun \& Zahn 2011, and references therein). Another plausible explanation for 
the confinement of the tachocline was that the Maxwell stress of the 
dynamo generated fields can provide a strong anisotropic angular momentum transport in the 
horizontal direction. For this 
mechanism to work, the dynamo-generated oscillatory magnetic field must penetrate the tachocline, which 
requires the value of the turbulent diffusivity to be $\ge 10^9$~cm$^2$s$^{-1}$. 
This is known as the {\it fast tachocline} mechanism (Forgacs-Dajka \& Petrovay 2001, 2002; Forgacs-Dajka 2003). 

In the fast tachocline scenario, the thickness of the tachocline depends on the magnetic field 
in a nonlinear way. On the other hand, the thickness of the tachocline is an important input parameter of 
flux transport dynamo models. Therefore it is expected to affect the dynamo solution in an unexpected way.
It is not even a priori clear whether the fast tachocline scenario and the \ftdm\ are compatible 
at all. The objective of the present work is to couple the simple feedback formula capturing the
essential physics of the \ftlm\ in a \ftdm\ and to see its response. Details can be found in Karak \& Petrovay (2013).

\section{Formula relating the tachocline thickness and the magnetic fields}
Following Forg\'acs-Dajka \& Petrovay (2001) the approximate relation between 
the mean (cycle averaged) tachocline thickness and Maxwell stress can be written as

\begin{equation} d_t^2 = \frac{C'\eta_t}{\bar B_p(t) \bar B(t)}  
  \label{eq:fits2new}
\end{equation}
where $\eta_t$ is the mean diffusivity in the tachocline, 
$\bar B$ and $\bar B_p$ ($= \sqrt{B_r^2 + B_\theta^2}$) are the means value of the 
toroidal and the poloidal field in the tachocline defined in the following way.

$\bar B (t) = \frac 2{\pi} \int_0^{\pi/2} \bar B (\theta,t)   \,\mathrm{d}\theta$ and 
$\bar B_p (t) = \frac 2{\pi} \int_0^{\pi/2} \bar B_p (\theta,t)  \,\mathrm{d}\theta$\\
while  $\bar B(\theta, t)$ and $\bar B_p(\theta, t)$ are the local radially
averaged values
of the toroidal and the poloidal field calculated as

$\bar B (\theta, t) = \frac 1{2d_0} \int_{r_t-d_0}^{r_t+d_0} |B  (r,\theta, t)|\,\mathrm{d}r$ 
with $d_0=0.015R$, and 

$\bar B_p (\theta, t) = \frac 1{2d_0} \int_{r_t-d_0}^{r_t+d_0} \bar B_p (r,\theta, t)\,\mathrm{d}r$
with $\bar B_p = \sqrt{(B_r^2 + B_\theta^2)}$\\

Note that the above relation (\ref{eq:fits2new}) is strictly valid for the average (temporal and the latitudinal) 
values of $B_p$ and $B$, not for the actual values at a given point in space and time. However it is 
worthwhile to use the above simple physically motivated relation to explore its effect in the dynamo
model.
\section{Results}

\begin{figure}[!h]
\begin{center}
\includegraphics[width=4.0in]{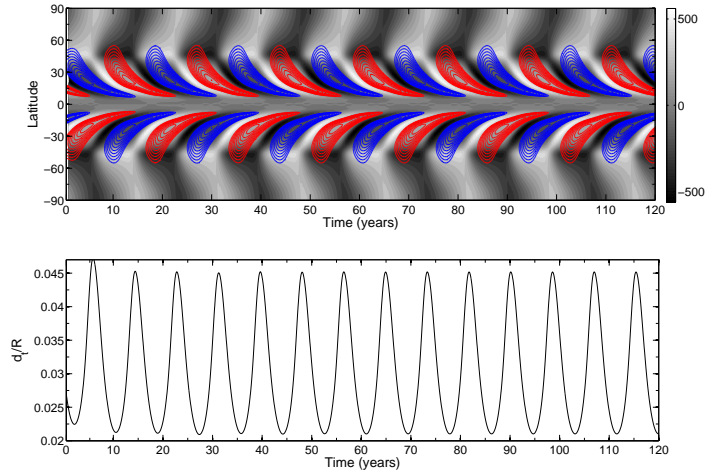}
\caption{Upper panel: butterfly diagram of magnetic fields for variable
tachocline thickness $d_t$ as given by Eq.~(\ref{eq:fits2new}). Contours
show the butterfly diagram of the toroidal field in the tachocline. 
Blue contours correspond to
positive toroidal field whereas red contours correspond to negative. 
The greyscale background shows the weak diffuse radial
field on the solar surface. Lower
panel: variation of tachocline thickness $d_t$ with time.}
\label{maxt}
\end{center}
\end{figure}

We use above relation for the tachocline thickness in a flux transport dynamo model 
(see Choudhuri 2011 for recent review). For the dynamo calculations we use the {\it Surya} code 
(Chatterjee, Nandy \& Choudhuri 2004) with modified parameters presented in Karak \& Petrovay (2013).
Usually the \ftdm\ uses fixed value of the \tl\ thickness. Here, however, we
consider a variable tachocline 
thickness based on Eq.~\ref{eq:fits2new}. The result for this calculation is shown in 
Figure \ref{maxt}. It is interesting to note is that this produces a stable solar-like dynamo solution. 
In addition, it produces a variable tachocline; the cyclic variation of \tl\
thickness is 
somewhat larger (up to $\pm30\%$) than the observational limit (e.g., Antia \& Basu 2011).

\begin{figure}[!h]
\begin{center}
\includegraphics[width=4in]{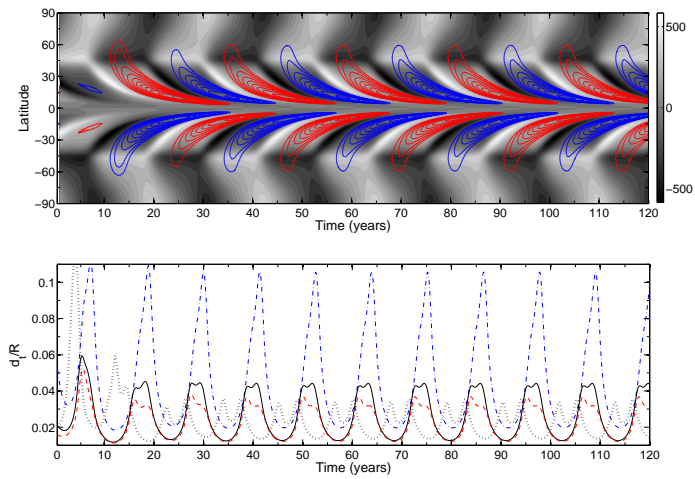}
\caption{Same as Figure \ref{maxt} but here the tachocline thickness [$d_t$] 
is determined by Eq.~(\ref{eq:fits2new2}). In the lower panel, 
the dash-dotted, solid, dashed and dotted lines are the values of $d_t$ 
at $75^0$, $60^0$,  $45^0$, and $15^0$ latitudes, respectively.}
\label{maxttheta}
\end{center}
\end{figure}

Next to explore the latitude dependence of tachocline thickness we write,
\begin{equation} d_t^2 = \frac{C'\eta_t}{\bar B_p(\theta, t) 
  \bar B(\theta, t)}.  
    \label{eq:fits2new2}
    \end{equation}
Figure \ref{maxttheta} shows the result. Note this procedure exibits a significant 
 latitudinal variation in the \tl\ thickness which agrees with the observations.

\begin{figure}[!h]
    \begin{center}
\includegraphics[width=5.0in]{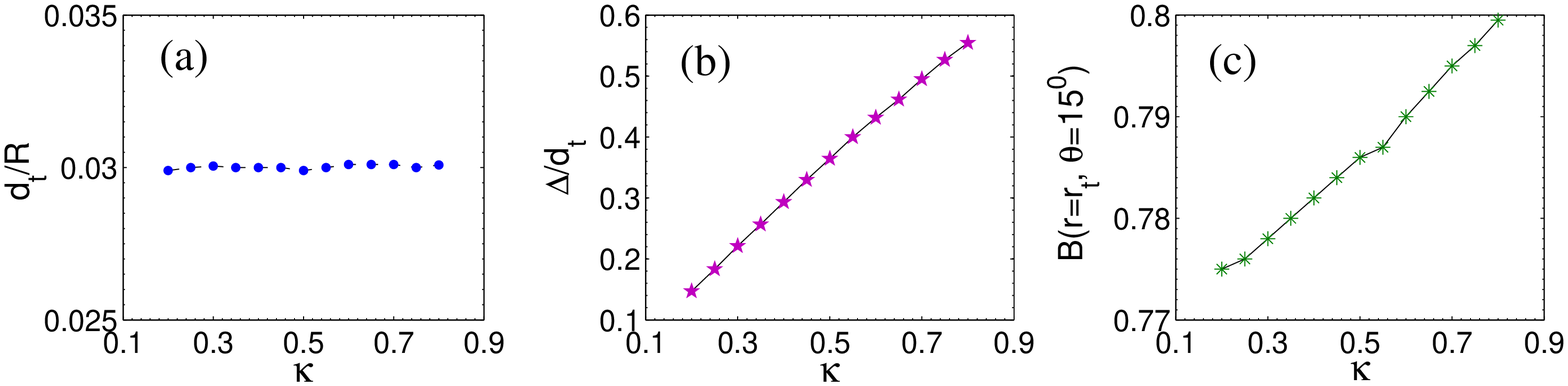}
    \caption{Variation of tachocline thickness [$d_t$] (a); of the amplitude
    variation of the tachocline thickness [$\Delta/d_t$] (b); and of toroidal
    field strength in the tachocline at $15^\circ$ latitude (c) as functions of $\kappa$ (the
    exponent in Eq.~\ref{eq:fitskappa}.} 
    \label{exponent}
    \end{center}
\end{figure}

In order to explore the sensitivity of our quantitative results to details 
of the feedback formula, we generalize Eq.~\ref{eq:fits2new2} in the following way,
\begin{equation} d_t = \frac{(C'\eta_t)^{1/2}}{[\bar B_p(\theta, t) 
  \bar B(\theta, t)]^{\kappa}}.  
  \label{eq:fitskappa}
\end{equation}
Recall that earlier we had $\kappa=0.5$. Although other values of $\kappa$ have no clear physical
meaning, this offers a way to test the robustness of our results. We repeated our 
calculations with different values of $\kappa$ from $0.2$ to $0.8$. In every run, we 
set $C'$ to fix the mean value of $d_t$ at around $0.03R$. Figure~\ref{exponent} shows the results.
We see that the amplitude variation of $d_t$ increases with $\kappa$.

\section{Conclusion}
We coupled the simple feedback formula of the \ftlm, relating the Maxwell stress of the 
dynamo generated magnetic fields with the tachocline thickness, into a \ftdm\ which is successful in explaining many
important aspects of the solar cycle (Choudhuri, Sch\"ussler \& Dikpati 1995; Dikpati \& 
Charbonneau 1999; Choudhuri \& Karak 2009; Karak 2011; Karak \& Choudhuri 2011, 2012, 2013;
Choudhuri \& Karak 2012; Karak \& Nandy 2012).
We find that the dynamo model is robust against the nonlinearity introduced in this way.
It produces a stable solar-like solution with a significant variation in the tachocline
with latitude and time. The thickness of the \tl\ varies from $0.02 R$ to $0.1 R$ as we 
move from low to high latitudes which is in agreement with the observations 
(e.g., Antia \& Basu 2011). However the solar cycle variation of tachocline thickness is quite significant, 
and somewhat higher than what the observational constraints suggest.
\begin{acknowledgements}
This work was supported by the Hungarian Science Research Fund (OTKA grants no.\
K83133 and K81421).
BBK thanks Department of Science and Technology, Government of
India for providing the travel support to participate this symposium.
\end{acknowledgements}


\begin{thebibliography}{}
\bibitem[Antia and Basu 2011]{ab11}Antia, H.M., Basu, S.: 2011, \textit{ApJ Lett.} 735, L45

\bibitem[Brun and Zahn 2006]{brun06}Brun, A.S., Zahn, J.-P.: 2006, \textit{A\&A} 457, 665

\bibitem[Charbonneau et al.\ 1999]{charbonneau99}Charbonneau, P. et al.\ 1999, \textit{ApJ}, 527, 445

\bibitem[Chatterjee, Nandy \& Choudhuri (2004)]{chatterjee}
Chatterjee, P., Nandy, D. \& Choudhuri, A.\ R. 2004,
\textit{A\&A}, 427, 1019

\bibitem[Choudhuri 2011]{ch11}
  Choudhuri, A.\ R. 2011, {\it Pramana}, 77, 77

\bibitem[Choudhuri \& Karak (2009)]{karak}
{Choudhuri, A. R., \& Karak, B. B.} 2009,
\textit{RAA}, 9, 953

\bibitem[Choudhuri \& Karak (2012)]{choudhurikarak}
{Choudhuri, A.\ R. \& Karak, B.\ B.} 2012,
\textit{Phys. Rev. Lett.}, 109, 171103

\bibitem[Choudhuri, Sch\"ussler \& Dikpati (1995)]{chou95}
{Choudhuri, A. R., Sch\"ussler, M., \& Dikpati, M.} 1995,
\textit{A\&A}, 303, L29

\bibitem[Dikpati \& Charbonneau (1999)]{dikpati99}
{Dikpati, M., \& Charbonneau, P.} 1999,
\textit{ApJ}, 518, 508

\bibitem[Forg\'acs-Dajka and Petrovay 2001]{petrovay01}Forg\'acs-Dajka, E., Petrovay, K. 2001, \textit{Solar Phys.} 203, 195
\bibitem[Forg\'acs-Dajka and Petrovay 2002]{petrovay02}Forg\'acs-Dajka, E., Petrovay, K. 2002, \textit{A\&A} 389, 629
\bibitem[Forg\'acs-Dajka 2003]{dajka03}Forg\'acs-Dajka, E. 2003, \textit{A\&A} 413, 1143
\bibitem[Garaud 2002]{garaud02}Garaud, P. 2002, \textit{MNRAS} 329, 1.


\bibitem[Karak (2010)]{karak10}
{Karak, B. B.} 2010,
\textit{ApJ}, 724, 1021

\bibitem[Karak \& Choudhuri (2011)]{karaknew}
{Karak, B.\ B., \& Choudhuri, A.\ R.} 2011,
\textit{MNRAS}, 410, 1503

\bibitem[Karak \& Choudhuri (2012)]{karak12}
{Karak, B.\ B., \& Choudhuri, A.\ R.} 2012,
\textit{Solar Phys.}, 278, 137

\bibitem[Karak \& Choudhuri (2013)]{karak13}
{Karak, B.\ B., \& Choudhuri, A.\ R.} 2013,
\textit{RAA}, arXiv:1306.5438

\bibitem[Karak \& Petrovay (2013)]{karakpetrovay}
{Karak, B.\ B., \& Petrovay, K.} 2013,
\textit{Solar Phys.}, 282, 321

\bibitem[Karak \& Nandy (2012)]{karaknandy}
{Karak, B.\ B., \& Nandy, D.} 2012, 
{\it ApJ Lett.}, 761, L13

\bibitem[Rudiger \& Kitchatinov 1997]{slowtach1}Rudiger, G., \& Kitchatinov, L.L. 1997, \textit{Astron. Nachr.} 318, 273

\bibitem[Spiegel and Zahn 1992]{spiegel92}Spiegel, E.A., \& Zahn, J.-P. 1992, \textit{A\&A} 265, 106
\bibitem[Strugarek, Brun, and Zahn 2011]{strugarek11}Strugarek, A., Brun, A.S. \& Zahn, J.-P. 2011, \textit{A\&A} 532, 34

\end{thebibliography}
\end{document}